# Meissner effect in honeycomb arrays of multi-walled carbon nanotubes

N.Murata[1,4], J.Haruyama[1,4, *], Y.Ueda[1], M.Matsudaira[1], H.Karino[1], Y.Yagi[4], E.Einarsson[2], S.Chiashi[2], S.Maruyama[2], T.Sugai[3,4], N.Kishi[3], H..Shinohara[3,4]

[1]*Aoyama Gakuin University, 5-10-1 Fuchinobe, Sagamihara, Kanagawa 229-8558 Japan*
[2]*Tokyo University, 7-3-1 Hongo, Bunkyo-ku, Tokyo 113-0033 Japan*
[3]*Nagoya University, Furo-cho, Chigusa, Nagoya 464-8602 Japan*
[4]*JST-CREST, 4-1-8 Hon-machi, Kawaguchi, Saitama 332-0012 Japan*
*Correspondence author; J.Haruyama (J-haru@ee.aoyama.ac.jp)

**We report Meissner effect for type-II superconductors with a maximum $T_c$ of 19 K, which is the highest value among those in new-carbon related superconductors, found in the honeycomb arrays of multi-walled CNTs (MWNTs). Drastic reduction of ferromagnetic catalyst and efficient growth of MWNTs by deoxidization of catalyst make the finding possible. The weak magnetic anisotropy, superconductive coherence length (~ 7 nm), and disappearance of the Meissner effect after dissolving array structure indicate that the graphite structure of an MWNT and those intertube coupling in the honeycomb array are dominant factors for the mechanism.**

New carbon-based superconductors, such as calcium-intercalated graphite with a transition temperature ($T_c$) of 11.5 K (*1,2*) and highly boron-doped diamond with $T_c$ = 4 K (*3*), have been recently found and attracted considerable attention, because a small mass of carbon may lead to high-$T_c$ superconductivity. Superconductivity in carbon nanotubes (CNTs), which are typical nano-carbon materials, has also attracted increasing attention (*4–7*). Three groups have experimentally reported superconductivity in different kinds of CNTs as follows; 1. with a $T_c$ as low as 0.4 K for resistance drops ($T_{cR}$) in ropes of single-walled CNTs (SWNTs) (*4*)(*6*), 2. with a maximum $T_{cR}$ as high as 12 K for an abrupt resistance drop in arrays of our multi-walled CNTs (MWNTs) entirely end-bonded by gold electrode (*7*), and 3. with a $T_c$ of 15 K for magnetization drops ($T_{cH}$) in thin SWNTs with diameters as small as 0.4 nm (*5*). However, no groups could report the observation of both the Meissner effect and the resistance drop down to 0 Ω in their respective systems (*21*), although these are indispensable factor to identify the occurrence of superconductivity in any materials. Moreover, it is extremely interesting to reveal how shielding current for Meissner effect or vortex occur and behave in one-dimensional space of CNTs.

One of main reasons for difficulty in observation of Meissner effect in CNTs is that most high-quality CNTs have been synthesized using ferromagnetic catalysts (e.g., Fe, Co, Ni) in previous studies. Such catalysts remain in the CNTs even after synthesis in some cases and destroy superconductivity. Hence, reduction of amount of the catalysts remaining after the synthesis of CNTs is crucial for observation of superconductivity.

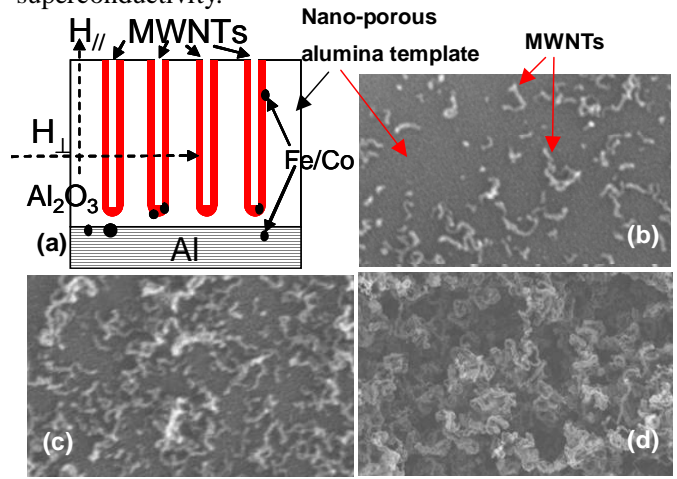

**Figure 1:** **(a)** Schematic cross-sectional view of the sample, which consist of aluminum substrate, nano-porous alumina membrane, MWNTs in the nano-pores, and Fe/Co catalyst remaining in some parts. **(b) – (d)** : SEM top view images of nano-porous alumina templates with array of MWNTs, which were synthesized using the smallest volume of Fe/Co catalyst **(b)** without deoxidization, **(c)** with deoxidization at 550 °C for 1 hour and **(d)** for 4 hours right before CVD process.

In the present study, we have synthesized honeycomb arrays of high-quality MWNTs in nanopores of alumina template ($Al_2O_3$) by chemical vapor deposition (CVD) using an iron (Fe)/cobalt (Co) catalyst and methanol (ethanol) gas, as shown in Fig.1(a) (7). Fe/Co catalyst was electro-chemically deposited into the bottom ends of nano-pores. Here, in our array of MWNTs, superconductivity for resistance drops could be observed as reported in our previous work (7), if even only one MWNT does not include such a ferromagnetic catalyst, because all supercurent can flow through this MWNT. In contrast, for magnetization measurements, even small amount of Fe/Co catalysts remaining in only a MWNT has obstructed observation of Meissner diamagnetism of the array. Therefore, in the present experiment, the conditions for the abovementioned electrochemical deposition of Fe/Co catalyst have been very carefully optimized for temperature of solutions at 40 ~ 62 °C, deposition voltages of 8 ~ 12 V, and time for 3 ~ 20 s in order to detect high $T_{cH}$. Consequently, the best condition, which allows the smallest volume of Fe/Co catalyst, was determined as the temperature of 40 °C, voltage of 8V, and time for 3 s.

However, it was found that amount of MWNTs synthesized in a porous alumina template was very small under this condition. Figure 1(b) shows a SEM top view image of alumina template right after synthesis of the array of MWNTs using this condition. Excess MWNTs, which grow from nano-pores on template surface, is very small volume in this figure, because of the very small amount of Fe/Co catalyst. This implies that some pores are even empty. In order to synthesize much larger volume of MWNTs under this condition, we have investigated many kinds of synthesis processes. As a result, we have successfully found that deoxidization of Fe/Co catalyst by ($H_2$ + Ar) gas right before CVD process is the most effective (22). Indeed, as shown in Fig.1(c) and (d), change of deoxidization time from 1 hour to 4 hours at 550 °C right before CVD drastically increased the volume of excess MWNTs and, hence, total volume of the synthesized MWNTs without empty pores as compared with Fig.1(b). We measured resistance of these samples and performed magnetization measurements only in the samples, which showed resistance drops due to superconductivity, using Superconducting Quantum Interference Device (Quantum Design).

The inset of Fig.2 (a) shows the magnetization, $M_\perp(T, H)$, as functions of the temperature (T) and the magnetic fields (H) applied perpendicular to the longitudinal tube axis ($H_\perp$) in the sample, which showed a small resistance drop due to superconductivity at $T_{cR}$ = 3 K (7), in the zero-field cooled (ZFC) regime As the field increases, the magnetization within positive values monotonically increases over the entire temperature range. This is attributed to very small volume of Fe/Co catalyst still remaining in some parts of the sample as shown in Fig.1(a). Moreover, such small-volume Fe/Co catalyst also obstructed detection of the Meissner effect in the FC regime.

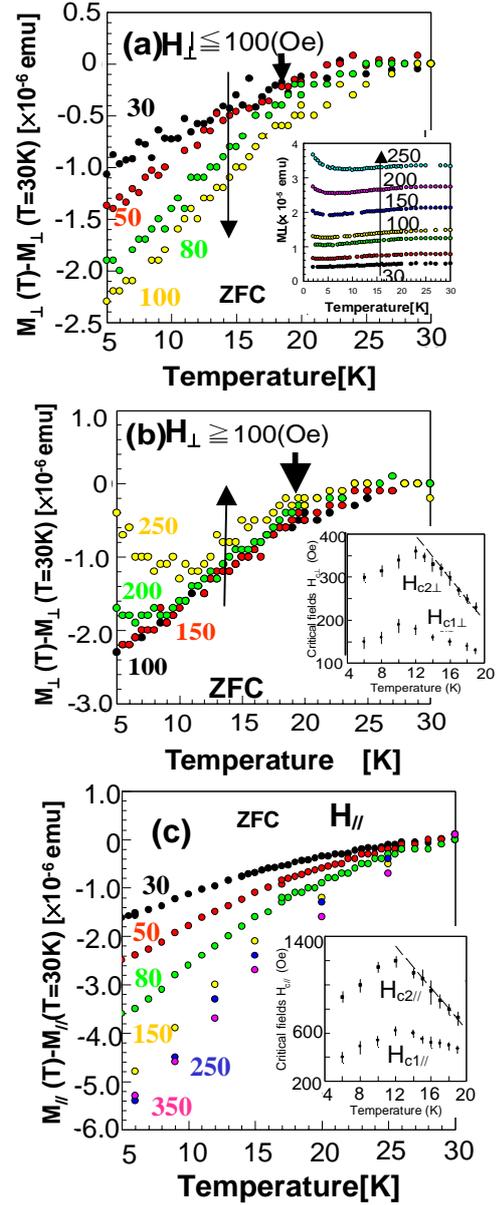

**Figure 2:** Normalized magnetization, $M_{n\perp//} = M_{\perp//}(T, H) - M_{\perp //}(T = 30 K, H)$, in the ZFC regime as functions of the temperature (T) and the magnetic fields (H). The number on each curve denotes the magnetic field (in Oe). The measured sample showed a small resistance drop with $T_c$ = 3 K. **(a), (b):** Results for H < 100 Oe and H > 100 Oe, respectively, when the fields are applied perpendicular to the longitudinal tube axis ($H_\perp$). Arrows mean the $T_{cH}$ of 19K. **(c):** Results for fields applied parallel to the tube axis ($H_{//}$). **Insets: (a)** Magnetization $M_\perp(T, H)$ of the sample used for main panel. **(b), (c)** Lower and upper critical fields (denoted by $H_{c1}$ and $H_{c2}$, respectively) as a function of T for fields applied **(b)** perpendicular ($H_{c\perp}$) and **(c)** parallel ($H_{c//}$) to the tube axis. These critical fields are determined from each $M_{n\perp//}(H) - H$ relationship at different temperatures.

On the other hand, the main panels of Figs.2 (a) and (b) show the normalized magnetization, $M_{n\perp} = M_\perp(T,$

H) – $M_\perp$ (T = 30 K, H), obtained from the inset of Fig.2 (a). As the temperature decreases, very evident magnetization drop can be observed from T = 19 K at H = 30 Oe in Fig.2 (a). The amplitude of this drop increases monotonically with an increase in fields up to 100 Oe over the entire temperature range below $T_{cH}$ = 19 K. Because no magnetization change is observable above T = 19 K at H = 30 and 50 Oe, this $T_c$ value is very distinct. In Fig.2 (b), the magnetization drop saturates at H = ~ 100 Oe and $M_{n\perp}$ increases with increasing fields above 100 Oe over the entire temperature range.

Importantly, this gradual and unsaturated magnetization drop below $T_c$ = 19 K occurs only in the arrays of MWNTs that exhibit the resistance drops (i.e., superconductivity) reported in ref.(*7*). This suggests a possibility that the magnetization drop is strongly associated with Meissner diamagnetization. As mentioned above, $M_{n\perp}(H)$ decreases below H = ~100 Oe and increases above it. These behaviors are in good agreement with those of type II superconductors (*9*). This evidently supports that the magnetization drops shown in the main panels of Figs. 2 (a) and (b) are attributed to the Meissner effect.

Figure 2 (c) shows the $M_{n//}$ (H) for fields applied parallel to the longitudinal tube axis in the ZFC regime. Although magnetization drops are also evident, the value of $T_{cH}$ = 19 K interestingly becomes unclear in contrast to that in Fig.2 (a). Because the field is applied parallel to the longitudinal tube axis, which corresponds to the application of fields within the graphene planes, the diamagnetism of graphite (*8*) that exists above T = 19K in any samples (even from room temperature in the case of some samples) becomes more significant than that in Fig.2 (a) (> 50 Oe) even at low fields. In Fig.2 (c), very gradual magnetization drop due to this diamagnetism smears out presence of evident $T_{cH}$ as mentioned above.

In Fig.2 (c), the drop saturates at H = ~ 350 Oe, which is considerably greater than H = ~ 100 Oe, while $M_{n//}(H)$ decreases above H = ~ 350 Oe. The observations support the fact that the present MWNTs are type II superconductors.

The insets of Figs.2 (b) and (c) show the lower and upper critical fields (denoted by $H_{c1}$ and $H_{c2}$, respectively) as a function of temperature for fields applied (b) perpendicular ($H_{c\perp}$) and (c) parallel ($H_{c//}$) to the tube axis. The temperature dependence of $H_c$'s at T > 12 K, at which $H_{c1}$ and $H_{c2}$ monotonically increase with a decrease in temperatures, and the relationship of $H_{c\perp} < H_{c//}$ are in qualitatively good agreement with that reported for type II superconductors within the Bardeen-Cooper-Schrieffer theory (*9*) and confirmed in $C_6Ca$ (*1,2*), $C_6Yb$ (*1*) and boron-doped diamonds (*3*).

In contrast, the values of $H_c$'s unconventionally decrease below T = 12 K due to the significant increase in the $M_{n\perp//}$ (H) with increasing fields at low temperatures. This is owing to Fe/Co catalyst still remaining in some part of the sample as well as the positive magnetization values shown in the inset of Fig.2(a). Moreover, the values of $H_c$'s do not become zero even around $T_{cH}$ = 19 K unconventionally, due to the diamagnetism of the graphite structure of the MWNTs above T = 19 K

Consequently, we conclude that the magnetization drops observed in Fig.2 are attributed to Meissner effect for type II superconductors. However, the $T_{cH}$ values observed in the samples that exhibited both the resistance and magnetization drops are greater than their $T_{cR}$ values in any samples. Indeed, for Fig.2-sample, $T_{cH}$ = 19 K was much larger than $T_{cR}$ = 3 K. The highest $T_{cH}$ value of 19 K shown in Fig.2 (a) is also greater than the highest $T_{cR}$ value of 12 K (*7*) between different samples. Although this seems to contradict to the conventional superconductivity in which $T_{cH}$ coincides with $T_{cR}$ (*9*), it can be understood as follows.

We reported $T_{cR}$ = 12 K for an abrupt resistance drop by end-bonding the arrays of MWNTs by Au electrodes in the same system as the present case (*7*), although the volume of Fe/Co catalyst was not largely reduced in it. $T_{cR}$ value was enhanced by an increase in the number of shells with current flows (*N*) in the MWNT, which was controlled by the type of the Au/MWNT junctions. The highest $T_{cR}$ of 12 K appeared only in the entirely end-bonded MWNTs with *N* = 9. In contrast, partially end-bonded MWNTs with 1 < *N* < 9 exhibited only slight and unsaturated resistance drops with low $T_{cR}$ values (the so-called signs of superconductivity). In the present magnetization measurement, we have measured the samples showing the signs of superconductivity. The sample for Fig.2 actually exhibited $T_{cR}$ = 3K for the sign of superconductivity.

The abovementioned contradiction is because the measured individual MWNT arrays contained more than $10^4$ MWNTs, and, hence, the MWNTs exhibiting $T_{cR}$ can differ from those exhibiting $T_{cH}$ in any arrays. The number of entirely end-bonded MWNTs ( ~ 100 as estimated in ref.(7)) showing $T_{cR}$ is significantly less than that of MWNTs exhibiting $T_{cH}$ (the number > 9900) due to difficulty in formation of the entire end-bonding. Hence, there is a high possibility that the MWNTs exhibiting $T_{cH}$ can have higher quality and a large number of metallic shells than the MWNTs exhibiting $T_{cR}$. Unlike resistance measurements, magnetization measurements require no electrodes and only the applied magnetic field can drive the shielding current for the Meissner effect in any such MWNTs of the array. Therefore, $T_{cH}$ can be higher than $T_{cR}$. The highest $T_{cH}$ = 19 K can be even greater than the highest $T_{cR}$ = 12 K In this sense, the $T_{cH}$ = 19 K represents the highest and intrinsic $T_c$ value for the MWNTs in the present arrays. It is evident that in the MWNT arrays, which contain MWNTs with resistance drops, favorable conditions can simultaneously exist for producing the Meissner effect in the other MWNTs within the same array.

Moreover, even the same individual MWNTs that are partially end-bonded can show $T_{cH}$ values greater than $T_{cR}$ values. Because the large value of $N$ led to intershell coupling and the suppression of the Tomonaga-Luttinger liquid (TLL) phase, the superconductive phase could easily overcome the TLL phase only in the entirely end-bonded MWNTs, resulting in the highest $T_{cR}$ of 12 K. However, the applied fields can drive the shielding current for the Meissner effect in all the shells even in partially end-bonded MWNTs, thereby resulting in the suppression of the TLL states, as in the case of the entirely end-bonded MWNTs. Hence, it becomes possible to exhibit the intrinsic $T_c$ ($T_{cH}$) value of the MWNTs, which is potentially greater than $T_{cR}$.

Here, we discuss the mechanisms of the observed Meissner effect. From the insets of Figs.2 (b) and (c), the values of $H_{c2\perp}(T = 0)$ = ~ 350 Oe and $H_{c2//}(T = 0)$ = ~ 1100 Oe can be estimated using $H_{c2\perp//}(T = 0)$ = $-0.69(dH_{c2}/dT|_{Tc})T_c$, subtracting the values of $H_{c1}$ and $H_{c2}$ at $T_c$ = 19 K that remain due to the diamagnetism of the graphite structure. These values are smaller than those of $C_6Ca$ and $C_6Yb$ (1, 2) as well as those at T = 2K. The Ginzburg-Landau (G-L) superconductive coherence length $\xi = [\Phi_0/2\pi H_{c2}(T = 0)]^{1/2}$, where $\Phi_0 = h/2e$ is the quantum of the magnetic flux, can be estimated from these $H_{c2\perp//}$ as $\xi_\perp$ = ~ 4.2 nm and $\xi_{//}$ = ~ 7.4 nm. On the other hand, the penetration length of the magnetic field $\lambda = (m^*/\mu n_s e^2)^{1/2}$ is estimated as order of ~100 nm. This $\lambda$ value is significantly larger than $\xi_\perp$ = ~ 4.2 nm and $\xi_{//}$ = ~ 7.4 nm. This result apparently supports the fact that the present MWNTs are type II superconductors. Furthermore, the values of $\xi_\perp$ = ~ 4.2 nm and $\xi_{//}$ = ~ 7.4 nm are smaller than those of $\xi_{\perp ab}$ = 13 nm and $\xi_{//ab}$ = 35 nm in $C_6Ca$ (1, 2) and $\xi$ = 10 nm in boron-doped diamond (3).

These small values of $\xi$ clarify the dominant mechanism of the observed superconductivity. At least, the following two origins of superconductivity in CNTs have been theoretically reported to date; 1. Contribution of the graphite structure (i.e., contribution of large values of $N$ as mentioned earlier) in the case of ropes of SWNTs (10) and the present MWNTs (7) and 2. Contribution of curvature (i.e., formation of $sp^3$ hybrid orbtals and enhancement of electron - phonon interactions) in thin SWNTs (11). Indeed, some of the present MWNTs include such a very thin SWNT as the innermost shell.

However, the values of $\xi_\perp$ = ~ 4.2 nm and $\xi_{//}$ = ~ 7.4 nm estimated above evidently indicate that the contribution of thin innermost shell is not a dominant factor in the present case, because this implies that the path of the shielding current for the Meissner effect and the vortex cannot be confined to the innermost shell with a diameter of ~ 0.5 nm. Moreover, from the magnetic anisotropy measurements, we obtain $\Gamma(H_{c2}, T = 0) = [H_{c2//} = 1100$ Oe$/H_{c2\perp} = 350$ Oe$]$ = ~ 3.1. This indicates the presence of weak magnetic anisotropy with the 3D Fermi surface, as observed in $C_6Yb$ ($\Gamma(H_{c2}, T=2K)$ = ~ 2) and $C_6Ca$ ($\Gamma(H_{c2}, T=0K)$ = ~ 1.6) (1,2). This also strongly supports the above-mentioned argument, because the diameter of 0.5 nm and length of 600 nm of the innermost shell should result in considerably higher magnetic anisotropy. Indeed, the resistance and magnetization drops were observable even in the array of MWNTs with innermost shells of a diameter > 1 nm. Therefore, we conclude that the contribution of graphite structure should be the dominant factor. This result is consistent with observation of resistance drops in MWNTs with the large $N$ and relevant because a breathing mode in MWNTs has not been reported to date.

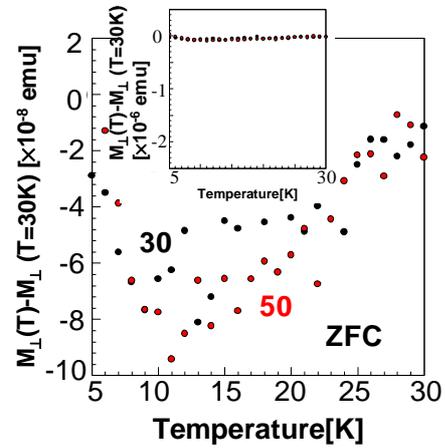

**Figure 3:** Normalized magnetization of MWNTs, which were placed on the Al substrate after dissolving alumina template, at H = 30 and 50 Oe in the ZFC. We dissolved the alumina template of the sample used for Fig.2 by using an ethanol solution. Subsequently, the entire solutions were spin-coated at random on a pure Al substrate without any losses and, then, the magnetization was measured by applying a field perpendicular to the Al substrate and tube. **Inset:** figure with the same scale as the main panel of Fig.2(a).

We also note that even individual MWNTs with an outer diameter of ~ 7 nm may be not sufficient to explain these results, because the geometric anisotropy of the outer tube diameter and tube length of 600 nm is still large as compared with $\Gamma(H_{c2}, T = 0) = \sim 3.1$ and the diameter can be roughly comparable to $\xi_\perp = \sim 4.2$ nm. From this standpoint, the magnetization measurements were performed on MWNTs, which were placed on the Al substrate at random after dissolving the $Al_2O_3$ template. Figure 3 shows the result and reveals the contribution of the intertube coupling in the array of MWNTs to the present Meissner effect. The magnetization shown in inset of Fig.3 is two orders of magnitude less than that shown in Fig.2 (a) over the entire temperature range. The magnetization drop observed below T = 19K in Fig.2(a) entirely disappears in Fig.3. Although only very slight magnetization drop is observable from very high temperatures, it's due to diamagnetism of graphite structure of MWNTs. This result implies a possibility that the intertube coupling in the honeycomb array of MWNTs strongly contributes to the present Meissner effect and that $T_{cH}$ of individual MWNTs is less than T = 5K. Shielding current or vortex exist crossing the MWNTs placed in the honeycomb array. This is just analogous to Abrikosov lattice for type II superconductor.

Because the mean spacing between neighboring MWNTs is ~ 5 nm and it is less than this value in some parts of the array, they are sufficient for the coupling of Cooper pair wave functions leaked by neighboring MWNTs or the tunneling of Cooper pairs. In fact, it has been reported that the intertube coupling in ropes of SWNTs leads to the following effects; 1. Screening of the electron - electron interaction (*12*) and 2. Increase in the density of states around $E_F$ apart from the energy gap (*13*). The intertube coupling observed in the present MWNT arrays can also lead to these effects, thereby enhancing $T_c$. This can be another reason for the present $T_{cH}$ as high as 19 K in addition to the contribution of the graphite structure (*14*).

Finally, we notice that the gradual and unsaturated magnetization drop shown in Fig.2 is in good correspondence with that in inhomogeneous superconductors such as $C_6Ca$ (*1*), $C_6S$ (*15*), and boron-doped silicon (*16*) and diamond (*3*). This indicates the possibility of inhomogeneous carrier doping into the present MWNT arrays. The weak magnetic anisotropy $\Gamma(H_{c2}, T = 0) = \sim 3.1$ also strongly supports this possibility. Moreover, all the shells in a MWNT should possess metallic behavior to obtain the $T_c$ ($> \sim 10K$) (*10*) as well as a high density of states around Fermi level. This can be easily achieved by high carrier doping.

Although we have used boron only to activate the chemical reaction for deposition of Fe/Co catalyst, excess boron atoms will be inhomogeneously included into the carbon network during the growth of the MWNTs, occasionally. In fact, it is reported that boron has been successfully doped into CNTs from catalyst with boron (*17, 18*). Some reports also have theoretically predicted the doping effects in CNTs on superconductivity (*19, 20*).

Further investigation is required to clarify the possibility of boron doping in the present MWNT arrays and to consistently understand all the properties of the superconductivity observed in the present study. The graphite structure and thin innermost shell in a MWNT and those intertube coupling in the array are expected to yield higher $T_c$ values like those in $C_{60}$ clusters and $MgB_2$, by combining with controlled high boron doping.

We are grateful to J. Gonzales, H. Bouchiat, C. Shoenenberger, M. Dresselhaus, D. Tomanek, C. Shtrunk, J. -P. Leburon, J. Xu, H. Fukuyama, J. Akimitsu, T. Ando, S. Saito, and R. Saito for their fruitful discussions and for providing encouragement during this study. We also thank J. Akimitsu, Y. Muranaka, Y. Oguro, and Y. Iye (ISSP, Tokyo University) for their valuable assistance in performing the SQUID measurements.

21. Recently, only the first group has reported drops in magnetization, which may indicate the occurrence of the Meissner effect (*6*).

22. In the present template, an aspect ratio of length/diameter of nano-pores is as high as ~ 90 = 700nm/8nm. Because it is very difficult to make efficient ethanol gas flow and reaction with Fe/Co surface at the bottom ends of such nano-pores, this deoxidization is very effective.